\documentclass[doublecol]{epl2} 

\usepackage{amsmath}
\usepackage{esint}
\usepackage{verbatim}

\newcommand{\beq}{\begin{equation}}
\newcommand{\eeq}{\end{equation}}
\newcommand{\be}{\begin{equation}}
\newcommand{\ee}{\end{equation}}

\newcommand{\avg}[1]{\left< #1 \right>} 
\newcommand{\de}{\upd}

\title{Joint statistics of quantum transport in chaotic cavities}
\shorttitle{Joint statistics of quantum transport in chaotic cavities} 

\author{Fabio Deelan Cunden\inst{1,2,3} \and Paolo Facchi\inst{4,2} \and Pierpaolo Vivo\inst{5}}
\shortauthor{F.D. Cunden, P. Facchi, P. Vivo}

\institute{                    
    \inst{1} School of Mathematics, University of Bristol, University Walk, Bristol BS8 1TW, United Kingdom\\
  \inst{2} Dipartimento di Matematica, Universit\`a di Bari, I-70125 Bari, Italy\\
  \inst{3} Istituto Nazionale di Fisica Nucleare (INFN), Sezione di Bari, I-70126 Bari, Italy\\
    \inst{4}Dipartimento di Fisica and MECENAS, Universit\`a di Bari, I-70126 Bari, Italy\\
    \inst{5} King's College London, Department of Mathematics, Strand, London WC2R 2LS,
United Kingdom
}


\abstract{
We study the joint statistics of conductance $G$ and shot noise $P$ in chaotic cavities supporting a large number $N$ of open electronic channels in the two attached leads. We determine the full phase diagram in the $(G,P)$ plane, employing a Coulomb gas technique on the joint density of transmission eigenvalues, as dictated by Random Matrix Theory. We find that in the region of typical fluctuations, conductance and shot noise are uncorrelated and jointly Gaussian, and away from it they fluctuate according to a different joint rate function in each phase of the $(G,P)$ plane. Different functional forms of the rate function in different regions emerge as a direct consequence of third order phase transitions in the associated Coulomb gas problem.}

\begin{document}

\maketitle

\section{Introduction}

We consider the problem of quantum electronic transport in mesoscopic devices. The typical setting is a cavity etched in semiconductors and connected to the external world by two attached leads. The cavity is brought out of equilibrium by an applied external voltage. Assuming that the average electron dwell time (spent inside the cavity) is
well in excess of the Ehrenfest time (when quantum effects kick in) at low temperatures and applied voltage, statistical properties of the electronic transport through a cavity exhibiting chaotic classical dynamics display remarkably universal features.

Random matrix theory (RMT) has provided tools and invaluable insights to successfully describe this universality
\cite{beenakker_transport}. In the RMT approach, the scattering process inside the cavity is governed by a scattering matrix $S$ drawn at random from the unitary group \cite{baranger_mello_94,jalabert_pichard_beenakker_94}.

The two incoming leads may in general support $N_1$ and $N_2$ (with $N_1\leq
N_2$) open electronic channels, i.e. different wave numbers of the incoming electronic plane waves. The $S$ matrix, having a natural block form $S= \left(\begin{matrix} r &
t^\prime
\\ t & r^\prime\end{matrix} \right), $ in terms of reflection $(r,r^\prime)$ and transmission $(t,t^\prime)$ matrices, connects the incoming and outgoing electronic wavefunctions. Conservation of electronic current implies that $S$ is unitary.
Landauer-B\"uttiker theory \cite{landauer_57,fisher_lee_81,butt2} expresses
most physical observables in terms of the eigenvalues $\bm{\lambda}=(\lambda_1,\ldots,\lambda_{N_1})$ of the
Hermitian matrix $t t^\dagger$. Unitarity of $S$ implies that $0\leq\lambda_i\leq 1$, $\forall i$. Hereafter, we will consider symmetric cavities $N=N_1=N_2$.

Important observables are the (specific, per channel) conductance and shot noise
\begin{equation}
G=\frac{1}{N}\sum_i{\lambda_i}, \qquad P=\frac{1}{N}\sum_i{\lambda_i (1-\lambda_i)}\ ,
\label{eq:GPdef}
\end{equation}
measured in units of the conductance quantum $G_0 = 2e^2/h$ and of $P_0 = 2e|\Delta V| G_0$ (with $\Delta V$ the applied voltage), respectively. They are expressed as \emph{linear statistics} on the transmission eigenvalues. Since the eigenvalues are $\lambda_i=\mathcal{O}(1)$, both $G$ and $P$ are intensive quantities of order $\mathcal{O}(1)$. More precisely, $0\leq G \leq 1$ and, since $\sum_i \lambda_i^2\leq (\sum_i \lambda_i)^2 \leq N \sum_i \lambda_i^2$, we have the useful geometric inequalities
\begin{equation}
0\leq P\leq G(1-G) \leq 1/4.
\end{equation}
The upper bound is attained at $G=1/2$, while the lower bound at $G=0$ or $G=1$.

Assuming now that $S$ is a random unitary matrix then implies that the $\lambda_j$s become (correlated) random
variables. What is their distribution? When the leads attached to the cavity are \emph{ideal} (no tunnel barriers), the appropriate choice is to assume $S$ \emph{uniformly} distributed in one of
Dyson's circular ensembles of random matrices, labeled by a parameter $\beta$:
$S$ is unitary and symmetric for $\beta=1$ (for systems that are invariant
under time-reversal), just unitary for $\beta=2$ (broken time-reversal invariance) and
unitary self-dual for $\beta=4$ (in case of anti-unitary time-reversal invariance). In this ideal
case the joint probability density function (jpdf) of transmission eigenvalues  is
given\cite{baranger_mello_94,jalabert1,forrester} by the Jacobi ensemble of RMT, namely
\begin{equation}
f(\bm{\lambda})=\frac{1}{Z_N} \prod_{i<j}{\left|\lambda_i-\lambda_j\right|^{\beta}}\prod_{i=1}^N\lambda_i^{\beta/2-1} ,  \quad \bm{\lambda}\in [0,1]^N,
\label{eq:jpdf}
\end{equation} where
$Z_N$ is a normalization constant enforcing $\int_{[0,1]^N}\de\bm\lambda f(\bm{\lambda})=1$. The random nature of $\bm{\lambda}$ promotes conductance and shot noise to random variables themselves, whose statistics is of paramount interest. The average and variance
of conductance and shot noise were considered, using perturbation theory in $1/N$, long
ago~\cite{baranger_mello_94,iida_weidenmuller_90,jalabert_pichard_beenakker_94}. In particular, as $N\to\infty$ the variance does not scale with $N$, as one would naively expects, but instead attains a
constant value $(\propto 1/\beta)$ depending only on the symmetry class, a phenomenon that has been dubbed
\emph{universal conductance fluctuations}.  This  value can be predicted from classical variance formulas \cite{beenakkervariance}. More recently, the classical theory of Selberg integral 
was employed and 
extended \cite{sommers_savin_06,savin_sommers_wieczorek_08,novaes_08,sommers_savin_schur} to address the 
calculation of transport statistics non-perturbatively (i.e. for a fixed and finite number of channels). The full
distribution of $G$ and $P$ is strongly non-Gaussian for few open channels, with power-law tails at the edge of their supports and non-analytic points in the interior
\cite{baranger_mello_99, sommers_savin_schur, kumar_pandey_10}. For finite $N$ and $\beta=2$, the Laplace transform of the full distribution was studied in \cite{kanzieper_ozipov_08} spotting a connection with integrable systems and Painlev\'e transcendents. The full distribution (including large deviation tails) for large $N$ was studied in \cite{vivo_majumdar_bohigas_08} using a Coulomb gas technique, where an error occurring in the asymptotic analysis in \cite{kanzieper_ozipov_08} was corrected. The statistics
of other observables was studied in \cite{novaes_07, novaes_08, vivo_vivo_08,
livan_vivo_11,texier}. The integrable theory of quantum transport in the ideal case,
pioneered in\cite{kanzieper_ozipov_08} for $\beta=2$, has been
recently completed including the other symmetry classes \cite{mezzadri_simm_11}. 

In this paper, we are concerned with the \emph{joint} statistics of conductance and shot noise~(\ref{eq:GPdef}) for a large number of open channels $N$ and in both regimes of small (typical) and large (atypical) fluctuations. Of particular interest are the jpdf of $G$ and $P$ 
\begin{equation}
\mathcal{P}(g,p)=\avg{\delta\left(g-G\right)\, \delta\left(p-P\right)}\ 
\label{eq:Pgpdef}
\end{equation} 
(where $\avg{\cdot}$ stands for the average with respect to the jpdf~\eqref{eq:jpdf}), and its Laplace transform 
\begin{align}
\hat{\mathcal{P}}(s,w) &=\big\langle e^{-\frac{\beta}{2} N^2(sG+wP)}\big\rangle.
\label{laplsw}
\end{align}

As discussed above, the arsenal of sophisticated techniques employed to tackle the calculation of joint cumulants for finite $N$, and their leading asymptotic behavior for $N\to\infty$, is truly impressive and virtually leaves no room for improvement in terms of mathematical rigor. In physical terms, however, it is  desirable to get a more intuitive understanding, based on simpler and more immediately decipherable formulas. Here we use the physically transparent Coulomb gas method to get  (at least to our eyes) a neater picture of the mutual relations between conductance and shot noise in such systems. The obtained formulas for the large deviation functions allow to recover effortlessly the leading terms of the joint cumulants. Additionally, the typical fluctuations of functions of both $G$ and $P$ can be easily investigated. As an example we will provide results on the \emph{Fano factor}
\begin{equation}
F=\frac{P}{G}
\label{eq:Fano}
\end{equation}
 which is essentially the ratio of the actual shot noise and the Poisson noise that would be measured if the system produced noise due to single independent electrons (for more details see the review~\cite{blanter}). Moreover, the \emph{full phase diagram} in the $(G,P)$ plane is obtained exactly and linked to the equilibrium electrostatic properties of the associated Coulomb gas of charged particles. 
We begin by first summarizing our results.

\section{Summary of results}

In this work we show, using a Coulomb gas technique that, for large $N$ and any $\beta>0$, the Laplace transform \eqref{laplsw} behaves as  \begin{equation}
\hat{\mathcal{P}}(s,w)\approx e^{-\frac{\beta}{2} N^2 J(s,w)},
\label{eq:GF}
\end{equation} 
where $J(s,w)$, independent of $\beta$, is the generating function (GF) of the joint cumulants of $G$ and $P$. 
Henceforth, the symbol $\approx$ stands for equivalence on a logarithmic scale. From this result, large deviation theory predicts that the jpdf of $(G,P)$ in~(\ref{eq:Pgpdef}) behaves asymptotically as 
\begin{equation}
\mathcal{P}(g,p)\approx e^{-\frac{\beta}{2} N^2 \Psi(g,p)},
\label{largedevrate}
\end{equation} 
where $\Psi(g,p)=\inf_{s,w}[J(s,w)-sx-wp]$ is the so-called \emph{joint rate function}. 
We find that $J(s,w)$ (and correspondingly $\Psi(g,p)$)  takes \emph{five} different functional forms in different regions of the $(s,w)$ (respectively, $(g,p)$) plane. This is a direct consequence of phase transitions in the associated Coulomb gas problem (see next Section). Across the lines of phase separation, the \emph{third} derivatives of the GF (the free energy of the associated Coulomb gas) are discontinuous, as it is typical in this type of problems (see \cite{majumdar_schehr_13} for a recent review). The rate function $\Psi(g,p)$ has a global minimum (zero) at $(g,p)=(1/2,1/8)$, corresponding to the average value of conductance and shot noise for large $N$, $\avg{G}=1/2$ and $\avg{P}=1/8$. Expanding the rate function around this minimum, we find that the \emph{typical} \emph{joint} fluctuations of conductance and shot noise are Gaussian, with a diagonal covariance matrix [see~Eq.~(\ref{ratef2})] implying the absence of cross-correlations to leading order. This is in agreement with earlier findings \cite{savin_sommers_wieczorek_08,mezzadri_simm_11,kanzieper_ozipov_08,sommers_savin_schur,novaes_07,cundenvivocovariance}. However, \emph{atypical} fluctuations far from the average are \emph{not} described by the Gaussian law, but rather by a different rate function. For $w=0$ ($s=0$) our GF reduces to the GF of conductance (shot noise) alone, computed in \cite{vivo_majumdar_bohigas_08}. In the next Section, we set up the Coulomb gas calculation.

\section{The  Coulomb gas} 

The suitable tool to perform a large dimensional analysis is the empirical density of transmission eigenvalues
\begin{equation}
\rho(\lambda)=N^{-1}\sum_i{\delta(\lambda-\lambda_i)}\ ,
\end{equation} 
a normalized random measure on the interval $[0,1]$.
The first step to compute the multiple integral \eqref{laplsw} is to realize that it can be written as
$\hat{\mathcal{P}}(s,w) =  Z_N(s,w)/Z_N(0,0)$, where 
\begin{equation}
Z_N(s,w)= \int\left[\mathcal{D}\rho\right]\, e^{-\frac{\beta}{2} N^2 E[\rho;s,w] },
\label{laplsw11}
\end{equation} 
and the \emph{energy density} functional
\begin{align}
\nonumber E& [\rho;s,w]=-\iint\log{\left|\lambda-\lambda^\prime\right|}\rho(\lambda)\rho(\lambda^\prime)\de\lambda\de\lambda^{\prime}\\
&+\int{V_{\mathrm{ext}}(\lambda;s,w)\rho(\lambda)\de\lambda} +\mathcal{O}(1/N)
\ , \label{action}
\end{align}
with $V_{\mathrm{ext}}(\lambda;s,w)=(s+w)\lambda-w\lambda^2$,
is defined on normalized spectral densities $\int{\rho(\lambda)\de\lambda}=1$.
Equation~\eqref{laplsw11} is the Gibbs-Boltzmann partition function of a system of charged particles confined on the interval $[0,1]$ of the real line, interacting via a (logarithmic) 2D Coulomb potential in an external single-particle potential $V_{\mathrm{ext}}$. The long-range, all-to-all nature of the interaction between the charges has the consequence that both energy contributions (interaction and external potential) are of order $\mathcal{O}(N^2)$. This is the origin of the $\mathcal{O}(N^2)$ speed of the large deviation functions \eqref{eq:GF} and \eqref{largedevrate}, in sharp contrast to the standard $\mathcal{O}(N)$ speed for classical large deviation theory of independent random variables~\cite{touchette}.
Using $\rho$, any linear statistics for large $N$ can be easily written down. In particular, 
\begin{equation}
G[\rho]=\int\lambda\, \rho(\lambda)\de\lambda, \;\; P[\rho]=\int\lambda(1-\lambda)\, \rho(\lambda) \de\lambda\ .\label{XYint}
\end{equation} 

In the large $N$ limit, the partition function~\eqref{laplsw11} 
is dominated by the saddle-point density
\begin{equation}
Z_N(s,w)\approx e^{-\frac{\beta}{2} N^2 E[\rho^\star;s,w]}.
\end{equation}
Here $\rho^\star(\lambda)$ (depending parametrically on $s$ and $w$) is the solution of the saddle point equation $\frac{\delta}{\delta\rho }E[\rho;s,w]
=0$, namely
\begin{equation}
\frac{\de}{\de\lambda}V_{\mathrm{ext}}(\lambda;s,w)=2\fint{\frac{\rho^\star(\lambda^\prime)}{\lambda-\lambda^\prime}\de\lambda^\prime}, \label{tricomi}
\end{equation}
for all $\lambda$ where the charge density exists $\rho^\star(\lambda)>0$, where $\fint$ denotes Cauchy's principal value. In terms of our electrostatic model, Eq. \eqref{tricomi} is the continuous version of the force balance condition for the charge cloud $\rho^\star(\lambda)$ to be in equilibrium. 

Originally due to Dyson \cite{Dyson}, this Coulomb gas technique with additional constraints has been developed  and used in many different problems~ \cite{Majumdar,vivo_majumdar_bohigas_07,majumdar_vergassola_09,depasquale,vivo_majumdar_bohigas_08,
vivopca,facchi,majumdar_schehr_13,marino,chen} which have apparently very little to do with each
other. 
Recently, an improvement of the method 
has been used to compute the joint statistics of two linear statistics in the Wishart-Laguerre ensemble of random matrices \cite{cundenvivospread,Grabsch}. We will adopt the same strategy below.

First, it is worth noticing a symmetry of this problem. The external potential is invariant (up to an immaterial constant) under the exchange $(s,\lambda)\to (-s,1-\lambda)$, namely,
\begin{equation}
V_{\mathrm{ext}}(1-\lambda; -s,w) = V_{\mathrm{ext}}(\lambda;s,w) - s.
\end{equation} 
Since the logarithmic repulsion is also invariant under the exchange $\lambda \to 1-\lambda$, the energy functional $E[\rho;s,w]$ inherits this symmetry. Therefore, the equilibrium spectral densities at $s$ and $-s$ are simply related by $\rho^{\star}(\lambda;-s,w)=\rho^{\star}(1-\lambda;s,w)$. Thus the phase diagram is invariant under the inversion $s\to -s$ and it is sufficient to study the problem for $s\geq0$. In the next Section, we compute the explicit solutions of the singular integral equation \eqref{tricomi} for any given value of $(s,w)$ and we find that \emph{five} different functional forms are possible for the saddle-point density $\rho^\star$ in various regions of the $(s,w)$ plane. 

\begin{figure*}[t]
\centering
\includegraphics[height=7cm]{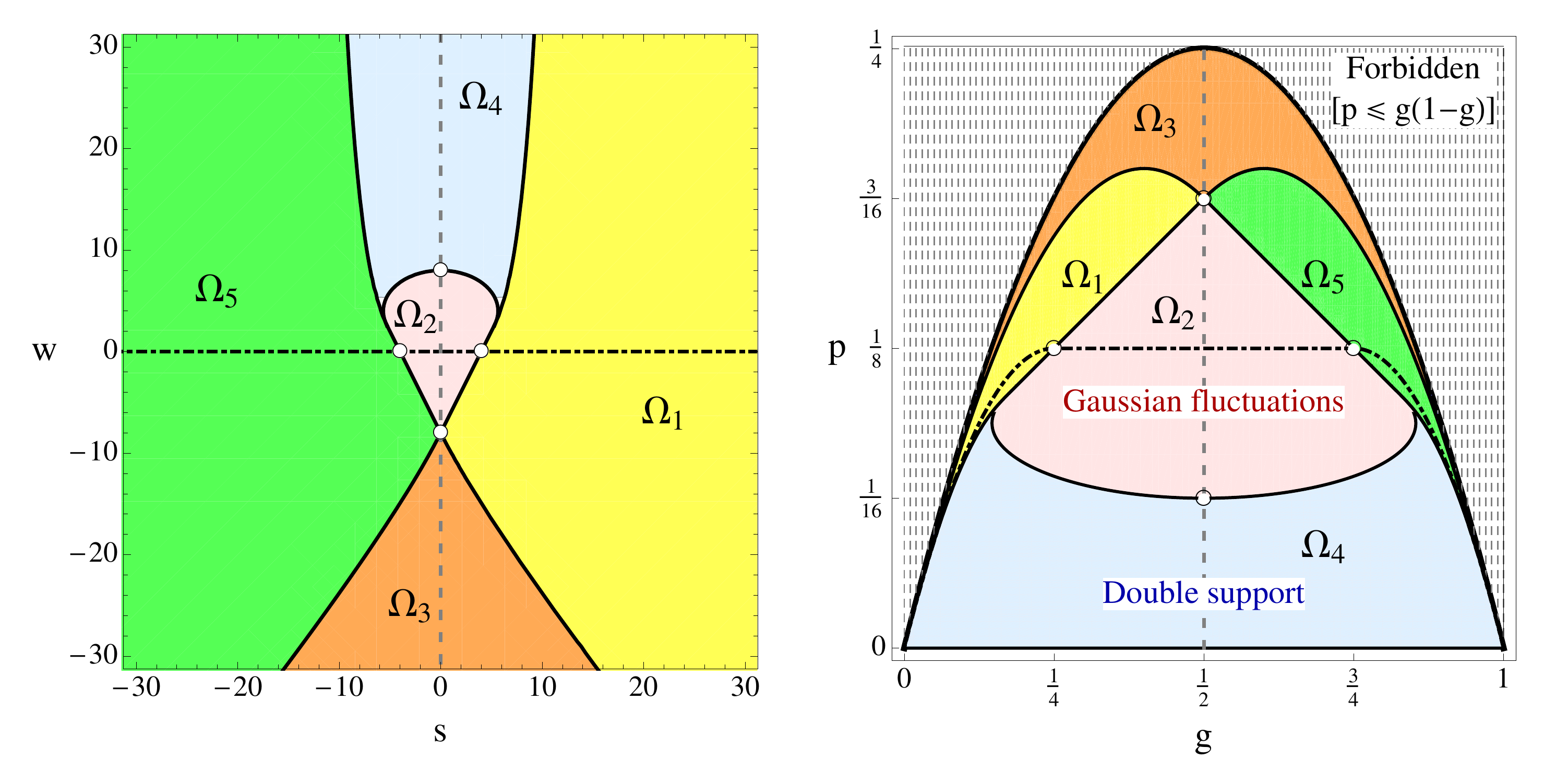}\quad\\
\vspace{.1cm}
\includegraphics[width=0.32\columnwidth]{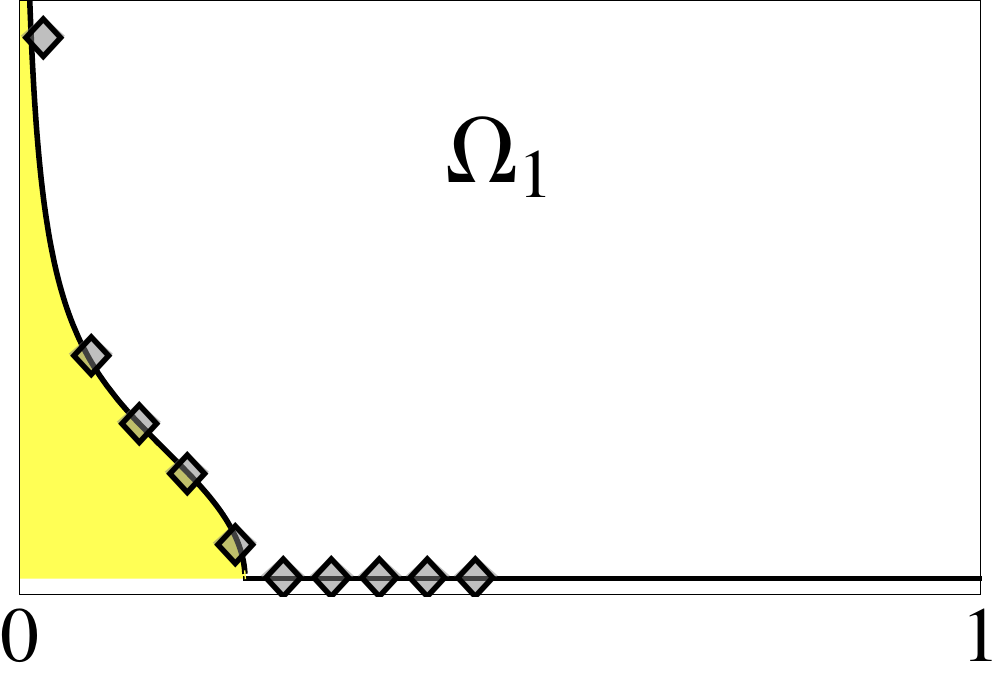}
\includegraphics[width=0.32\columnwidth]{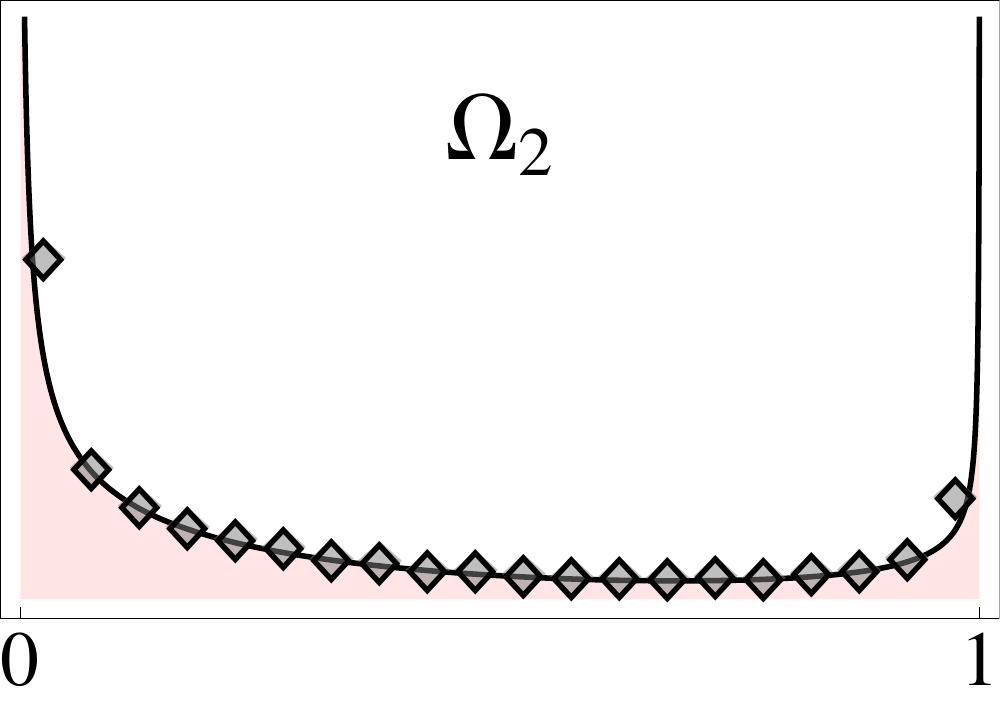}
\includegraphics[width=0.32\columnwidth]{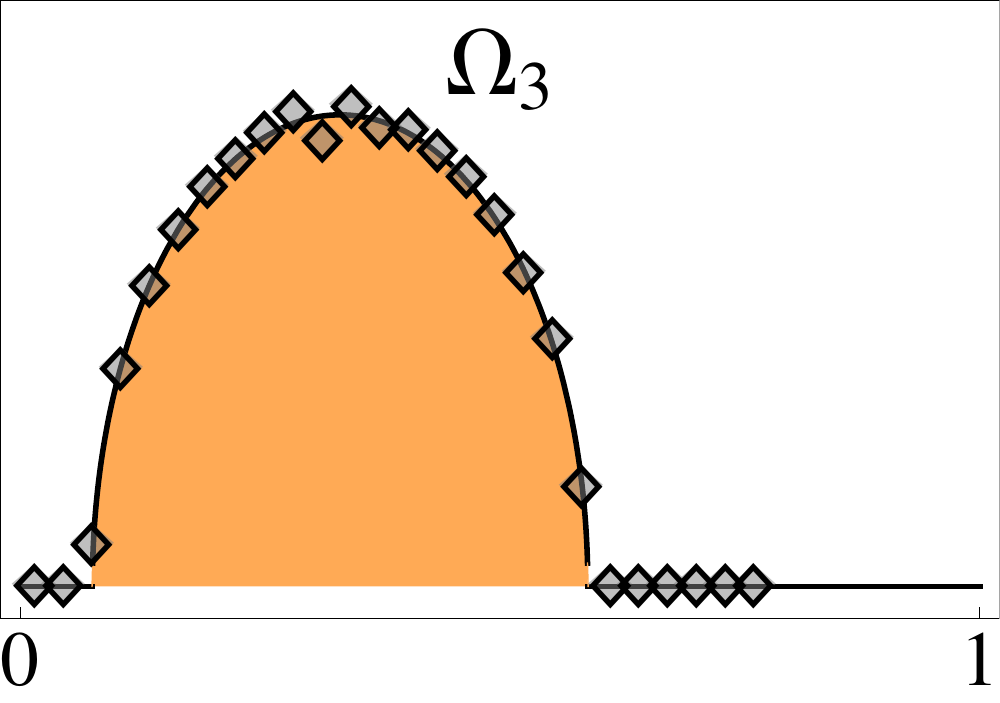}
\includegraphics[width=0.32\columnwidth]{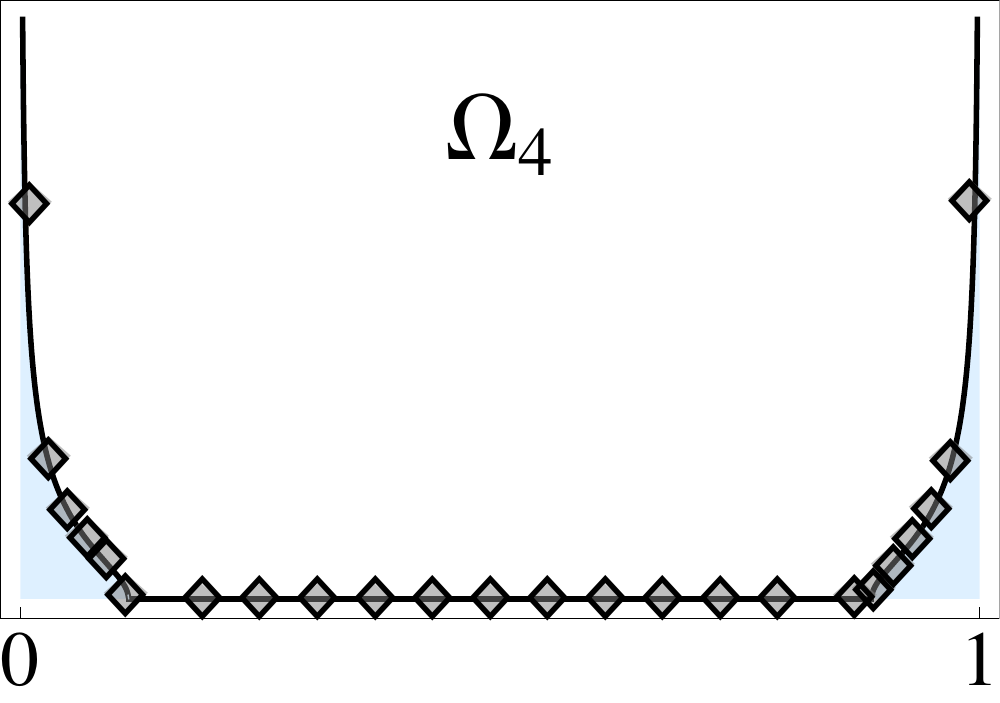}
\includegraphics[width=0.32\columnwidth]{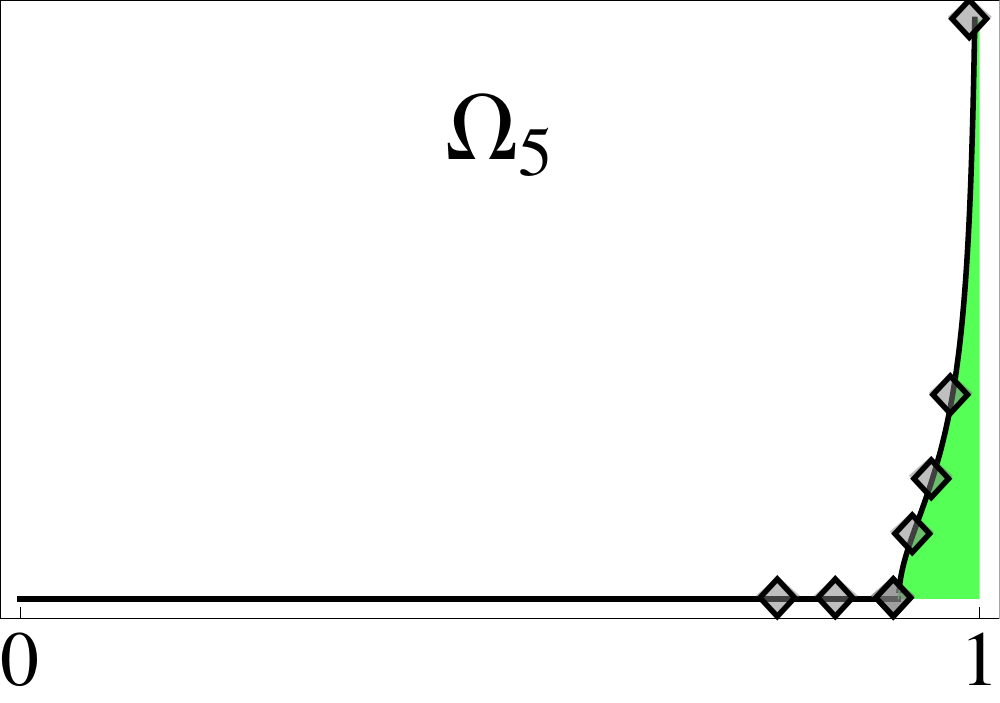}
\caption{\textbf{Top Left:} Phase diagram in the Laplace $(s,w)$ plane. The five domains $\Omega_i$, $i=1,\dots,5$ correspond to different spectral profiles $\rho^{\star}(\lambda)$ of the transmission eigenvalues~\eqref{rho1}-\eqref{rho5}. The critical lines that separate the phases are reported in Appendix in~\eqref{g13}-\eqref{g45}. Notice that the phases $\Omega_1$ ($\mathrm{supp}\rho^{\star}=[0,b]$)  and $\Omega_5$ ($\mathrm{supp}\rho^{\star}=[a,1]$) are separated by the phases $\Omega_2$, $\Omega_3$ and $\Omega_4$. The diagram is symmetric with respect to the line  $s=0$. The lines $w=0$ (black dotdashed) and $s=0$ (gray dashed) have been studied in \cite{vivo_majumdar_bohigas_08}. We recover their critical points $s_{\mathrm{cr}}=\pm4$ and $w_{\mathrm{cr}}=\pm8$. \textbf{Top Right:} The corresponding phase diagram in the conductance/shot noise $(g,p)$-plane. For completeness, the critical points and the lines corresponding to $s=0$, $w=0$ are also reported in real space.  \textbf{Bottom:} Different spectral profiles $\rho^{\star}(\lambda)$ of the Coulomb gas. The analytical curves are superimposed to points from a Montecarlo simulation
of a Coulomb gas of $N=30$ particles in the external potential $V_{\mathrm{ext}}(\lambda;s,w)$.}
\label{fig:phasediag}
\end{figure*}

\section{Phase diagram} 

We have identified five domains $\Omega_j$, $1\leq j\leq 5$ in the $(s,w)$-plane that correspond to five different phases of the Coulomb gas. The functional forms of $\rho^\star_j$ and their supports in the phase $(s,w)\in\Omega_j$ , with $j=1,\dots,5$, read
\begin{align}
\rho^{\star}_1(\lambda)&=\frac{q_1(\lambda)}{2\pi}\sqrt{\frac{b-\lambda}{\lambda}}, &&\lambda\in [0,b] \label{rho1}, \\
\rho^{\star}_2(\lambda)&=\frac{q_2(\lambda)}{2\pi}\frac{1}{\sqrt{\lambda(1-\lambda)}},&&\lambda\in [0,1],  \label{rho2}\\
\rho^{\star}_3(\lambda)&=\frac{q_3(\lambda)}{2\pi}\sqrt{(\lambda-a)(b-\lambda)},&&\lambda\in [a,b],  \label{rho3} \\
\rho^{\star}_4(\lambda)&=\frac{|q_4(\lambda)|}{2\pi}\sqrt{\frac{(a-\lambda)(\lambda-b)}{\lambda(1-\lambda)}} ,&&\lambda\in [0,a]\cup[b,1],\!\!  \label{rho4}\\
\rho^{\star}_5(\lambda)&=\frac{q_5(\lambda)}{2\pi}\sqrt{\frac{\lambda-a}{1-\lambda}},&&\lambda\in [a,1],  \label{rho5}
\end{align}
with $0<a<b<1$. Here $q_j(\lambda)$ are polynomials in $\lambda$, 
whose expressions, together with those of $a$ and $b$, depend on the values $(s,w)$ and are given explicitly
in~\eqref{eq:polys}-\eqref{eq:ends} of the Appendix. 
The five different domains and the corresponding densities are shown in Fig. \ref{fig:phasediag}.

The physical picture is quite intuitive in terms of the interplay between the logarithmic interaction between the charges and the external potential $V_{\mathrm{ext}}$. For moderate values of $s$ and $w$ (i.e. $(s,w)\in\Omega_2$) the external potential is too weak compared to the logarithmic repulsion between the charges, so that the gas invades the whole interval $[0,1]$. For sufficiently large positive values of $s$ ($\Omega_1$), the potential is strong enough to attract the charges towards $\lambda=0$ and confine the gas in $[0,b]\subset[0,1]$. In a similar way, large negative values of $s$ ($\Omega_5$) provide a repulsive potential that pushes the gas away from the origin towards the hard wall located at $\lambda=1$. When $w<0$ and $|s|<-w$, $V_{\mathrm{ext}}$ has a minimum that attracts the charges; if this well is sufficiently deep (i.e. $(s,w)\in\Omega_3$) the Coulomb gas is trapped in the minimum of the potential and does not feel the hard walls. On the other hand, when $w>0$ and $|s|<-w$ , $V_{\mathrm{ext}}$ has a maximum that repels the charges;  in $\Omega_4$ the peak is sufficiently high to cause the Coulomb gas to split into two separated components that tend to stay far apart.

The single-support solutions have been found employing a theorem due to Tricomi~\cite{Tricomi1} for the singular equation~\eqref{tricomi}. The theorem provides the general form of $\rho^{\star}(\lambda)=\rho^{\star}(\lambda;C,a,b)$ with a single support. It depends on three arbitrary constants that are determined by imposing the normalization condition $C=\int{\rho^{\star}(\lambda)\de \lambda}=1$ and the behavior of $\rho^{\star}$ at the two edges $a$ and $b$ of the support. This procedure provides a density supported on a single interval and parametrized by $s$ and $w$.

Multiple-support solutions of the saddle-point equations are more complicate and involve in general more arbitrary constants. For example for a solution $\rho^{\star}(\lambda)=\tilde{\rho}(\lambda;C,a,b,\kappa)$ with double support in $[0,a]\,\cup\,[b,1]$ (the only multiple support scenario that we have in our problem), we have to impose the overall normalization $C=1$, the regularity condition at the edges $\rho^{\star}(a)=\rho^{\star}(b)=0$ and an extra constraint, i.e. the so-called \emph{filling fraction} of one of the two component of the density,  $\int_0^a{\rho^{\star}(\lambda)\de \lambda}=\kappa$~\cite{Eynard}. How to fix this filling fraction? 
At equilibrium, the filling fraction $\kappa$ is such that a small variation $\delta\kappa$ does not change the energy at first order. This condition means that, at equilibrium, the work required to move a single charge from $a$ to $b$ is null, and can be easily proved to read\footnote{Note that the integral runs \emph{outside} the support of $\rho^\star$, therefore it needs to be interpreted as the integral of the analytic continuation of $\rho^\star(\lambda)$ to values of $\lambda$ not in the support.} $\int_a^b{\mathrm{i}\rho^{\star}(\lambda)\de \lambda}=0$ where $\mathrm{supp}\rho^{\star}=[0,a]\cup[b,1]$~\cite{Eynard,Jurkiewicz}. 

These kinds of constraints are often cumbersome to evaluate. However, if one is only interested in finding the critical line of phase transition between the single-support phases and the double-support phase in the $(s,w)$-plane, a practical strategy is the following. The condition of null work for moving a charge from $a$ to $b$ 
should be satisfied for any double-support solution. In the limit this will be valid also at the ``birth'' of the double-support phase when $b=1$. Then, the requirement $\int_{b^{\star}}^1{\rho^{\star}(\lambda)\de \lambda}=0$ with $\rho^{\star}(\lambda)$ supported on $[0,b^{\star}]$ provides a threshold $b^{\star}$ and therefore an equation connecting $s$ and $w$ at the ``birth-of-second-cut", i.e. precisely the line of phase transition in the $(s,w)$-plane.

For given values of $(s,w)$, the solution $\rho^\star$ provides the \emph{typical} configuration of eigenvalues yielding prescribed values for 
$G$ and $P$
from \eqref{XYint}. 
Both are functions of $s$ and $w$: 
\begin{equation}
\label{eq:gpdef}
g(s,w)=G[\rho^\star], \qquad p(s,w)=P[\rho^\star].
\end{equation} 
In particular, for $(s,w)=(0,0)$ (uncostrained problem) we obtain from \eqref{rho2}  and  \eqref{eq:polys} the arcsine law $\rho_{\mathrm{as}}(\lambda)=1/(\pi\sqrt{\lambda(1-\lambda)})$ providing the average values $\avg{G}=1/2$ and $\avg{P}=1/8$ from \eqref{XYint}. In general, \eqref{rho2} in $\Omega_2$ describes the typical fluctuations around the average values $\avg{G}$ and $\avg{P}$. Values of $(s,w)\in\Omega_1$ correspond to a configuration of the transmission eigenvalues \eqref{rho1} yielding smaller values of the conductance $g\ll\avg{G}$
while large deviations $g\gg\avg{G}$ correspond to $(s,w)\in\Omega_5$ and are described by~\eqref{rho5}. Similarly, large values for $P$ are given by a typical configuration of eigenvalues as in \eqref{rho3} in phase $\Omega_3$, while smaller values for $P$ in phase $\Omega_4$ are ascribed to the double support solution \eqref{rho4}. As the point $(s,w)$ moves in the Laplace plane, the configuration of eigenvalues $\rho^{\star}(\lambda)$ changes according to \eqref{rho1}-\eqref{rho5}. The transition across the critical line $\gamma_{ij}$ (the boundary of two  phases $\Omega_i$ and $\Omega_j$) corresponds to a change in shape of the Coulomb gas density, which signals a corresponding phase transition in the  quantities $G$ and $P$. In the next Section, we use the explicit functional forms of the density in various regions of the $(s,w)$ plane to show how the GF and the rate functions can be  computed.

\section{Joint large deviation functions} 
First, one inserts the equilibrium density $\rho^\star$ (whose explicit expressions are given in Eqs.~(\ref{rho1})-(\ref{rho5})) into \eqref{XYint} and computes the corresponding single integrals, yielding the order $\mathcal{O}(1)$ quantities $g(s,w)$ and $p(s,w)$. These are in turn related to the cumulant GF $J(s,w)$ in~\eqref{eq:GF} and the rate function $\Psi(g,p)$ in~\eqref{largedevrate} via the differential relations 
\begin{align}
\de J(s,w)&=g(s,w)\de s+p(s,w)\de w\ ,\label{dJ}\\
-\de \Psi(g,p)&=s(g,p)\de g+w(g,p)\de p \label{dPsi}\ ,
\end{align}
complemented with the condition $J(0,0)=0$ and $\Psi(g(0,0),p(0,0))=\Psi(1/2,1/8)=0$. Here $s(g,p)$ and $w(g,p)$ are the solution of $s=g(s,w)$ and $w=p(s,w)$. The expressions \eqref{dJ} and \eqref{dPsi} are known as Maxwell relations among  thermodynamic potentials, in our case the Helmholtz free energy and the enthalpy. The  application of  standard thermodynamics arguments to the Coulomb gas  can be rigorously justified by  means of large deviation principles \cite{ldp}.  In fact, relations  \eqref{dJ} and \eqref{dPsi} are \emph{equivalent} statements: one is in Laplace space -- the other one is in real space. By plugging~(\ref{rho1})-(\ref{rho5}) into (\ref{eq:gpdef}), one easily obtains the explicit expressions of $g(s,w)$ and $p(s,w)$ in each region of the phase space.
The  lines $w=0$ and $s=0$ (yielding the GF of conductance \emph{or} shot noise alone) were studied in \cite{vivo_majumdar_bohigas_08} and we easily recover the known results by taking the limit $w\to0$ (or $s\to 0$) in our expression. 
The joint cumulant GF $J(s,w)$ follows by a careful integration of the differential form \eqref{dJ} and the continuity requirement through the critical lines.
By starting from the center $(0,0)$ of region $\Omega_2$, where $J(0,0)=0$, and integrating, one   obtains 
\begin{equation}
J(s,w)=\frac{s}{2}-\frac{s^2}{32}+\frac{w}{8}-\frac{w^2}{256},\quad\mbox{if } (s,w)\in\Omega_2 \ .
\label{J2}
\end{equation}
Therefore, in the region of typical fluctuations, $G$ and $P$ are (to leading order) uncorrelated  Gaussian variables with average values $\avg{G}=1/2$ and $\avg{P}=1/8$, $\mathrm{Var}{G}=1/(8 \beta N^2)$, $\mathrm{Var}{P}=1/(64\beta N^2)$, and $\mathrm{Cov}{(G,P)}=0$:
\begin{equation}
\mathcal{P}(G,P)\approx e^{-4\beta N^2\left(G-\frac{1}{2}\right)^2-32\beta N^2\left(P-\frac{1}{8}\right)^2}\ .
\label{ratef2}
\end{equation}
Since the higher derivatives of $J(s,w)$ vanish at $(s,w)=(0,0)$ we also conclude that the higher order cumulants are $\kappa_{\ell,m}(G,P)=\mathcal{O}(1/N^{\ell+m-1})$ for $\ell+m>2$.

For generic $s$ and $w$, the integration of $\de J$ provides the cumulant GF:
\begin{equation}
J(s,w)= J(s_0,w_0)+\int_{(s_0,w_0)}^{(s,w)}\de J(s^\prime,w^\prime), 
\label{eq:Jswint}
\end{equation}
for $(s,w)\in \Omega_i$. Here,  one has to  choose a suitable initial point on the boundary with a phase $\Omega_j$ already computed, namely  $(s_0,w_0)\in \Omega_i\cap \Omega_j$. The expression~(\ref{eq:Jswint}) can be evaluated by integrating in the complex plane and applying the residue theorem. Across the lines of phase separation, one finds that the GF is not analytic, since its third derivatives are discontinuous.

As an additional bonus, from the joint limiting behavior of $G$ and $P$ we can deduce the  distribution of functions of both $G$ and $P$ like the Fano factor~(\ref{eq:Fano}).
We stress the fact that $F$ is \emph{not} a linear statistics of the transmission eigenvalues. From~\eqref{ratef2} we conclude that the typical fluctuations of $F$ are asymptotically Gaussian
\begin{equation}
\mathcal{P}(F)\approx e^{-\frac{16}{3}\beta N^2\left(F-\frac{1}{4}\right)^2} ,
\label{eq:Fano_distr_sym}
\end{equation}
with average value $\avg{F}=1/4$ and and variance $\mathrm{Var}{F}=3/(32\beta N^2)$.
To the best of our knowledge the asymptotic distribution~\eqref{eq:Fano_distr_sym} is a new result.

\section{Conclusions}
In summary, our analysis gives an overall picture of the joint statistics of conductance $G$ and shot noise $P$ for an ideal chaotic cavity supporting a large number $N$ of electronic channels in the two attached leads. We employed 
a Coulomb gas technique 
to establish the large deviation formulas \eqref{eq:GF} and \eqref{largedevrate}, governing the behavior of the joint cumulant GF and the joint rate function of $G$ and $P$. 
We were  able to obtain the full phase diagram in the $(s,w)$ and $(G,P)$ planes and we found that both $J(s,w)$ and $\Psi(g,p)$ acquire five different functional forms in different regions of their domain. These different expressions are a direct consequence of phase transitions in the associated Coulomb gas problem. Across the lines of phase separation, the third derivatives are discontinuous, implying a third order phase transition.
\acknowledgments
This work has been partially supported by ``Investissements d'Avenir" LabEx PALM (ANR-10-LABX-0039-PALM), EPSRC Grant No.\ EP/L010305/1 and ``Gruppo Nazionale di Fisica Matematica'' GNFM-INdAM. We thank Ricardo Marino for the initial collaboration on this topic and Eiji Kawasaki for helpful advices.

\section{Appendix}
Let us denote by $\gamma_{ij}$ the critical curve that separates the phases $\Omega_i$ and $\Omega_j$ in the $(s,w)$ plane. We report their explicit expressions
\begin{align}
\gamma_{13}\cup\gamma_{35}\, &: \;
|s|=-w-\sqrt{-8w} &\mbox{for }& w\leq -8\ ,\nonumber\\
\gamma_{12}\cup\gamma_{25}\, &:\; |s|=w/2+4&\mbox{for }& -8\leq w\leq 8/3\ ,\nonumber\\
\gamma_{24}\, &:\; s^2+2w^2-16w=0 &\mbox{for }&  8/3 \leq w\leq 8\ ,
\label{g13}
\end{align}
while the curves $\gamma_{14}\cup\gamma_{45}$ are the solutions of the equation

\begin{align}
& 2w\phi_1\left(b_1(|s|,w)\right) = [2(|s|+w)+\sqrt{(|s|+w)^2-24w}]
\nonumber\\ 
& \qquad\qquad\qquad\qquad    \times \phi_0\left(b_1(|s|,w)\right)/3\ , \qquad w\geq 4,\label{g14}\\
& \phi_{k}(b) =\mathrm{i}\int_b^1\de x\,x^k\sqrt{(b-x)/x}\ ,\quad 0<b<1\ .\label{g45}
\end{align}

The polynomials $q_i(\lambda)$ and the edges $a_i,b_i$ of the supports $\mathrm{supp}\rho^{\star}$ are given by 
\begin{align}
q_1(\lambda)&=[-6 w \lambda+2 (s+w)+\sqrt{(s+w)^2-24 w}]/3\ ,  \nonumber\\
q_2(\lambda)&=2w \lambda^2-(2w+s)\lambda+w/4+s/2+2\ , \nonumber\\
q_3(\lambda)&=-2w\ ,  \nonumber\\
q_4(\lambda)&=2(t-wx)\ ,  \nonumber\\
q_5(\lambda)&=q_1(1-\lambda;-s,w)\ , 
\label{eq:polys}
\end{align}
and 
\begin{align}
a_1&=0, \qquad b_1 =[s+w-\sqrt{(s+w)^2-24 w}]/(3 w), \nonumber\\
a_2&=0,  \qquad b_2=1,\nonumber\\
a_3 &=(s+w+\sqrt{-8w})/(2 w), \qquad b_3 =1-a_3(-s,w),\nonumber\\
a_4 &=[s+(2-b_4) w-2 t]/w,\nonumber\\
b_4&:\,\int_{a_4}^{b_4}2(t-wx)
\sqrt{(a_4-x)(x-b_4)}/\sqrt{x(1-x)} \de x=0,\nonumber\\
t&= \left( s+2 w- b_4 w\right)/3\nonumber\\
&-\sqrt{s^2+2 (2 b_4-1) s w-w \left(24-w-8 b_4 w+8 b_4^2 w\right)}/6,\nonumber\\
a_5&=1-b_1(-s,w), \qquad b_5=1,
\label{eq:ends}
\end{align}
respectively.

\end{document}